\documentclass[12pt]{article}
\usepackage{sublabel}
\usepackage{graphicx}
\usepackage{epsf}
\usepackage{wrapfig}
\usepackage{epsfig}\usepackage{graphicx,epsf}
\usepackage{latexsym}
\usepackage{amsmath}
\usepackage{amssymb}
\usepackage{exscale}
 \advance\oddsidemargin by -1in
 \advance\evensidemargin
 by -1in
 \marginparsep
 0.4cm
 \advance\topmargin by
 -0.0in
 \makeindex
\def\Vec#1{\mbox{\boldmath $#1$}}
\def\bra#1{\langle#1|}
\def\ket#1{|#1\rangle}

\def\beq{\begin{equation}}
\def\eeq{\end{equation}}

\def\beqy{\begin{eqnarray}}
\def\eeqy{\end{eqnarray}}
\def\ie{\textit{i.e.}}

\newcommand{\ber}{\begin{displaymath}}
\newcommand{\eer}{\end{displaymath}}
\newcommand{\bey}{\begin{eqnarray}}
\newcommand{\eey}{\end{eqnarray}}

\begin{document}
\title{\textbf{Effects of correlations on the total neutron-Nucleus cross section
at high energies}}
\author{\large M. Alvioli$^a$, C. Ciofi degli Atti$^a$, I. Marchino$^a$,
H. Morita$^{b}$ and V. Palli$^a$}
\maketitle
\begin{center}
  {$^a$ Department of Physics, University of Perugia, and
\\ INFN, Sezione di Perugia, via A. Pascoli, Perugia, I-06100, Italy\\
$^b$ Sapporo Gakuin University, Bunkyo-dai, Ebetsu, Hokkaido 069-8555, Japan}
\end{center}
\begin{abstract}
\noindent
The total neutron-Nucleus cross section has been calculated within an approach
which takes into account nucleon-nucleon correlations, Glauber multiple scattering
and inelastic shadowing corrections. Nuclear targets ranging from $^{4}He$ to $^{208}Pb$
and neutron incident momentum ranging from $3$ to $300$ $GeV/c$, have been considered.
Correlations have been introduced by two different approaches leading to the same
results. The commonly used approximation, consisting in treating nuclear effects only
by a product of one-body densities, is carefully analyzed and it is shown that the
effects of realistic correlations resulting from modern nucleon-nucleon interactions
and realistic correlations resulting from realistic nucleon-nucleon interactions and
microscopic  ground state calculation of nuclear properties cannot be disregarded.
\end{abstract}
%\date
\newpage
%\pacs{PACS}
The total neutron-Nucleus cross section at high energies has been
the object of many calculations for its dependence  is very sensitive
to various effects, such as Glauber elastic \cite{glauber}) and
Gribov  inelastic \cite{gribov}) diffractive shadowing, which are relevant for the
interpretation of color transparency phenomena \cite{jen01,boris1} and relativistic
heavy ion processes  \cite{boris2}. The major mechanism  governing the
total cross section is
Glauber inelastic shadowing, but  a quantitative  explanation of the experimental data
has been achieved in the past only by considering also the effects
of inelastic shadowing \cite{murthy,nikolaev}. All calculations so far performed
were based upon the  so called one body density approximation,
in which all terms but the first one of the correct  expansion of the square of the nuclear
wave function in terms of density matrices \cite{foldy} are disregarded, which amounts
to neglect all kinds of nucleon nucleon correlations.
The necessity to investigate the effects of correlations on the total cross section
$\sigma^{tot}_{nA}$ was pointed out by several authors \cite{jen01,nikolaev}.
It is precisely the  aim of this  work to present the results of calculations of the total
neutron-Nucleus cross section within an approach based upon realistic many-body correlated wave
functions \cite{alv01} obtained with realistic nucleon-nucleon interactions \cite{vuotto},
 Glauber multiple scattering and Gribov inelastic shadowing.

%-------------------------------------------------------------------------------SECTION
\section{Basic formalism}

\begin{figure}[!htp]
\centerline{\epsfig{file=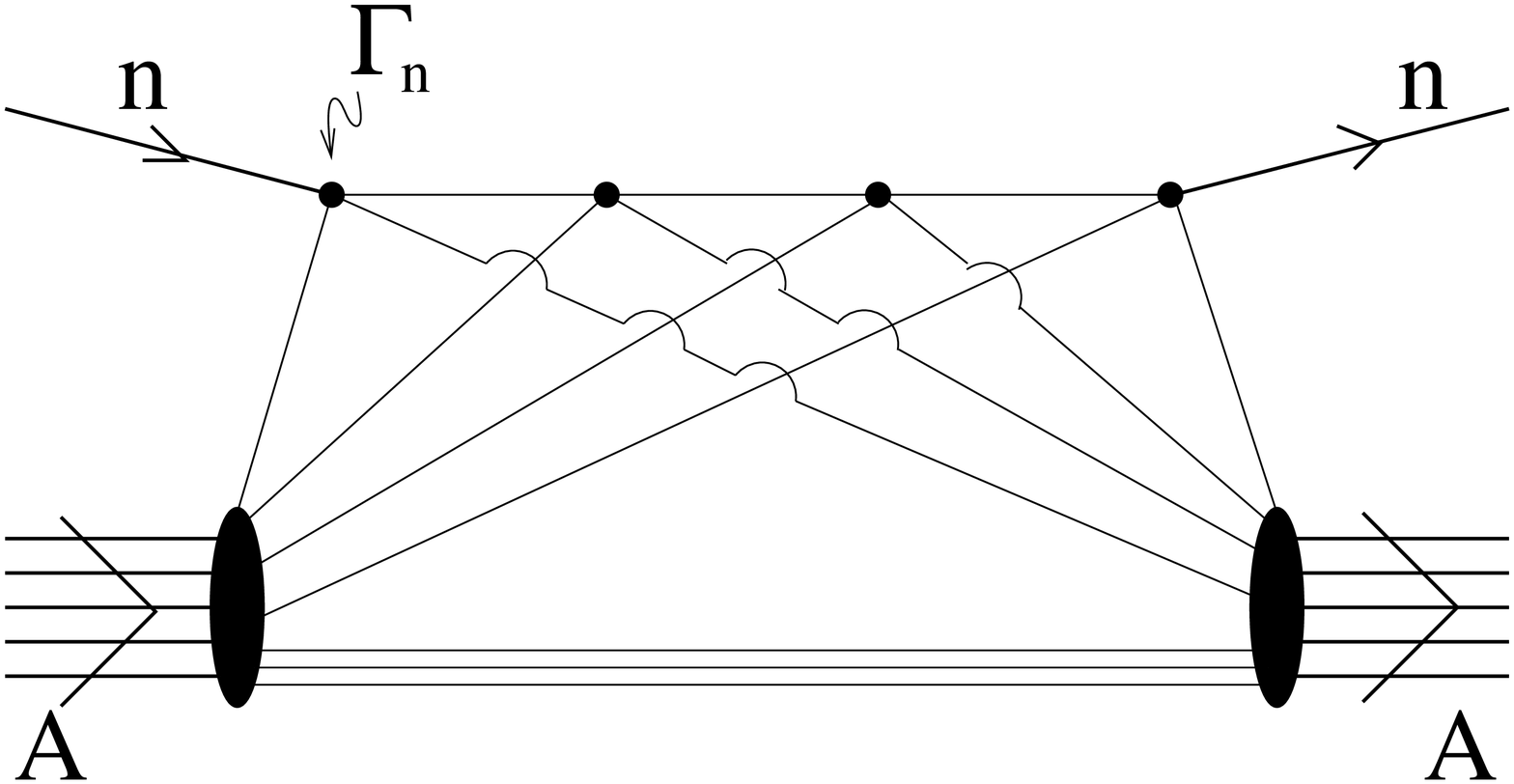,width=7.5cm}}
\centerline{\Large a)\normalsize}
\centerline{\epsfig{file=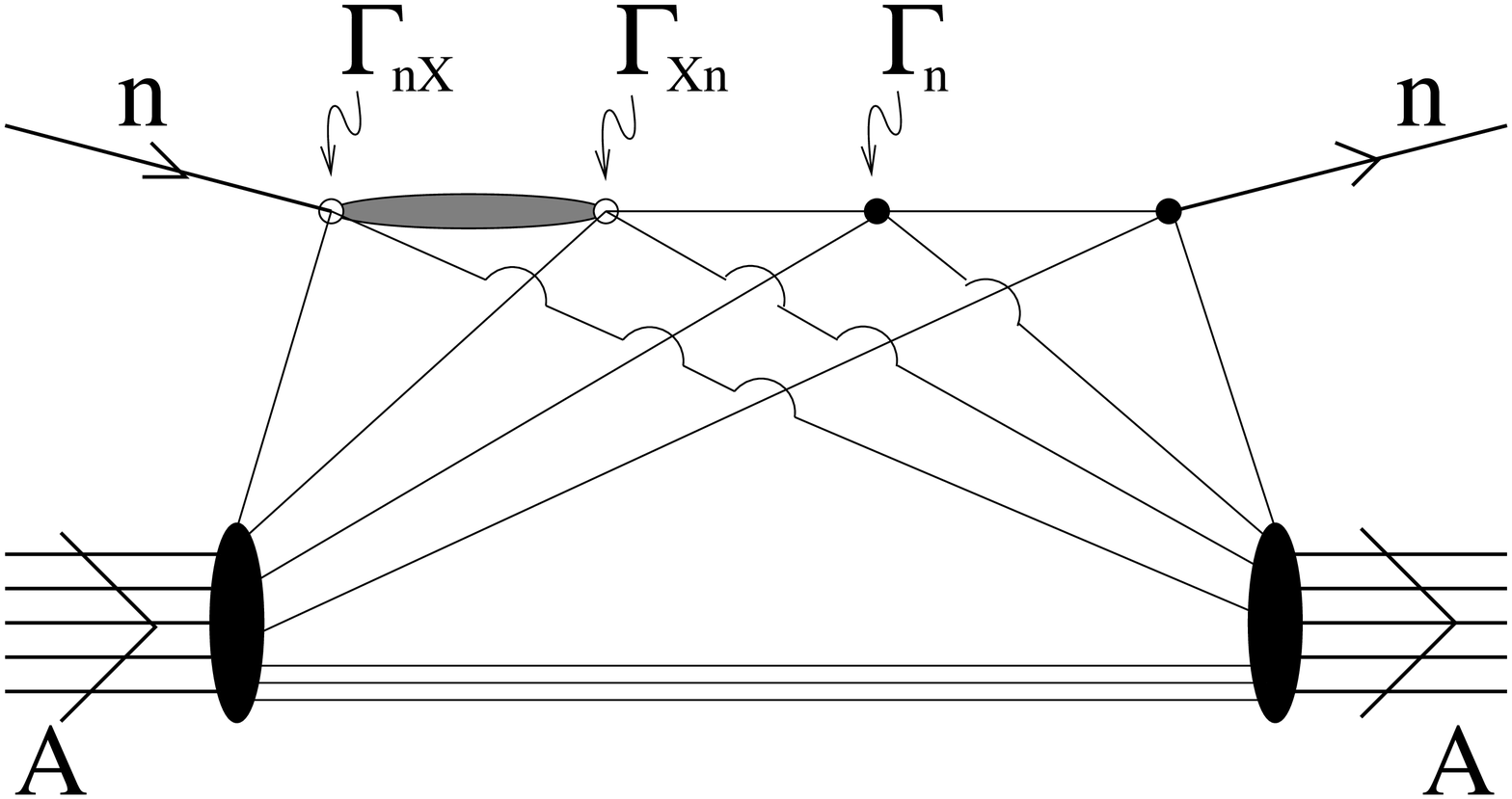,width=7.5cm}}
\centerline{\Large b)\normalsize}
\centerline{\epsfig{file=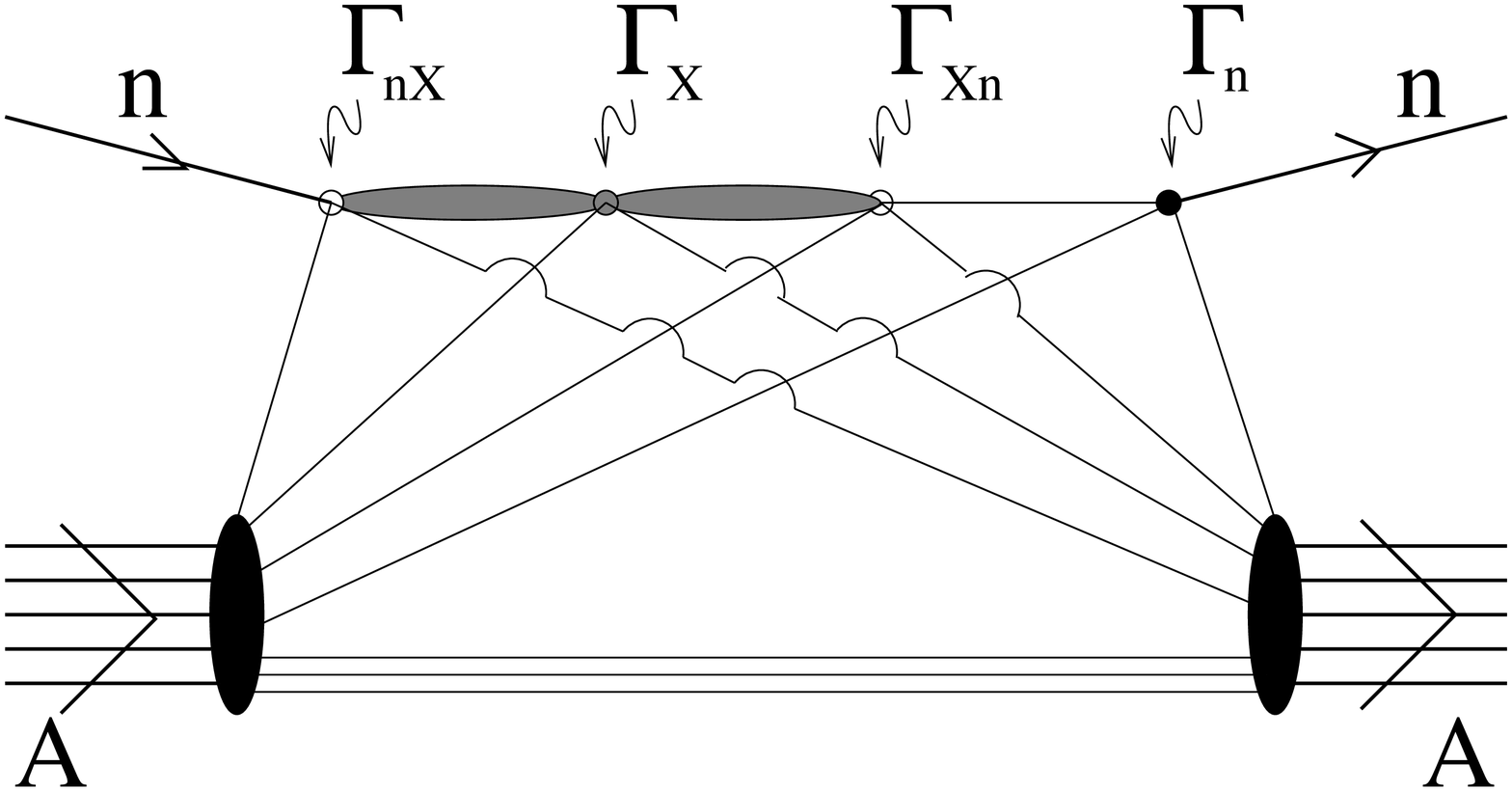,width=7.5cm}}
\centerline{\Large c)\normalsize}
\caption{Typical diagrams describing elastic neutron-nucleus scattering
at high energies.
\textit{a)} Glauber multiple scattering; \textit{b)} and \textit{c)} inelastic shadowing.}
\label{Fig1}
\end{figure}

Considering both Glauber (G) elastic scattering and Gribov
 inelastic  shadowing (IS) ,  the total cross section on a nucleus $A$
can be written as follows
\beq
\sigma_A^{tot}\,=\,\sigma_A^G\,+\,{\sigma}_A^{IS}\,=\,\frac{4\pi}{k}\,Im
\left[\,F_{00}^G(0)\,+\,F_{00}^{IS}(0)\,\right]
\eeq
where $\sigma_A^G$ and ${\sigma}_A^{IS}$ denote respectively
the  Glauber and inelastic shadowing contributions,  and  $F_{00}^G(0)$ and $F_{00}^{\Delta}(0)$
the corresponding forward elastic scattering amplitudes related to the full nuclear profile $\Gamma_{00}^{G(IS)}$
as follows
\beq
F_{00}^{G(IS)}(0)\,=\,\frac{ik}{2\pi}\,\int d\Vec{b}_n\,\Gamma_{00}^{G(IS)}(\Vec{b}_n)
\eeq
The Glauber nuclear profile describing the elastic scattering of the neutron has the usual form
\begin{equation}
\Gamma_{00}^G(\Vec{b}_n)\,=\,1\,-\,\prod_{j=1}^A\langle \psi_0
\left |\left[\,1\,-\,\Gamma_{N}(\Vec{b}_n-{\bf s}_j)\,\right]
\right | \psi_0\rangle\,,
\label{SG}
\end{equation}
where $\psi_0 \equiv \psi_0({\bf r}_1,{\bf r}_2,{\bf r}_3,...{\bf r}_A)$, with
${\bf r}_j=({\bf  s}_j,z_j)$, is  the ground state wave function of the target
nucleus, $\Vec{b}_n$  the impact parameter of the neutron moving along the $z$-axis,
and ${\mit\Gamma}_{N}(\Vec{b}_n)$
the nucleon-nucleon elastic  profile function. The Inelastic Shadowing Profile
should describe  the  diffractive dissociation of the neutron via
the process $n + N \rightarrow X + N$ and its  de-excitation to the ground state by
the process $X + N \rightarrow n + N$, as well as the elastic  scattering
of $X$ off the target nucleons. The three  processes are described  by the inelastic profiles
$\Gamma_{NX}$ and $\Gamma_{XN}$, and by the elastic  profile $\Gamma_X$, respectively. In
our approach, as in Ref. \cite{jen01}, we will consider
only two non-diagonal transitions, i.e.  $n + N \rightarrow X + N$ and
$X + N \rightarrow n + N$, and the elastic scattering of $X$. The corresponding diagrams
are shown in Fig. \ref{Fig1}. Within such an approximation, the
Inelastic Shadowing  profile can  be written in the following form \cite{jen01}:
\beqy
\Gamma_{00}^{IS}(\Vec{b}_n)&=&
\sum_{X}\left\{\bra{\psi_o}\sum_{i<j}^{A}\,\Gamma_{NX}(\Vec{b}_n-\Vec{b}_j)
\,\Gamma_{XN}(\Vec{b}_n-\Vec{b}_{i})\,e^{i\,q_{X}(z_i-z_j)}\Theta (z_j-z_i)
\,\times\right.\nonumber\\
&&\left.\times \prod_{k\neq i,j}^{A}[1-\Gamma_{X}(\Vec{b}_n-\Vec{b}_{k})]
\,\Theta(z_k-z_i)\,\Theta(z_j-z_k)\,\times\right.\nonumber\\
&&\left.\times\,\prod_{l\neq i,j}^{A}[1-\Gamma_{N}(\Vec{b}_n-\Vec{b}_{l})]
\,\Theta(z_i-z_l)\,\Theta(z_l-z_j)\ket{\psi_o}\right\}
\label{Fdelta}
\eeqy
where
\beq
q_{X}\,=\,k_{n}\,-\,k_{X}
\eeq
is the longitudinal momentum transfer. The basic nuclear ingredient appearing in
Eqs. (\ref{SG}) and (\ref{Fdelta}) is the square of the nuclear
wave function $\psi_0$, which can be written in terms of density matrices as follows
\cite{foldy}:
\beqy
\left|\,\psi_o(\Vec{r}_1,...,\Vec{r}_A)\,\right|^2&=&\prod_{j=1}^A\,\rho_1(\Vec{r}_j)
\,+\,\sum_{i<j}\,\Delta(\Vec{r}_i,\Vec{r}_j)\,\prod_{k\neq i,j}\rho_1(\Vec{r}_k)\,+\nonumber\\
&&\hspace{0.5cm}+\hspace{-0.2cm}\sum_{(i<j)\neq(k<l)}\hspace{-0.5cm}\Delta(\Vec{r}_i,\Vec{r}_j)
\,\Delta(\Vec{r}_k,\Vec{r}_l)\hspace{-0.2cm}\prod_{m\neq i,j,k,l}\hspace{-0.2cm}\rho_1(\Vec{r}_m)
\,+\,\dots\,;
\label{psiquadro}
\eeqy
in which $\rho_{1}(\Vec{r}_i)$ is the one-body density matrix
\beq
\rho_{1}(\Vec{r}_1)=\int \prod_{j=2}^A d\Vec{r}_j
\,\left|\psi_0(\Vec{r}_1,...,\Vec{r}_A)\right|^2 \delta\left(\sum\Vec{r}_j\right)
\label{onebody}
\eeq
and the \textit{two-body contraction}  $\Delta$ is defined as follows:
\beq
{\Delta(\Vec{r}_1,\Vec{r}_2)}\,=\,\rho_2(\Vec{r}_1,
\Vec{r}_2)\,-\,\rho_{1}(\Vec{r}_1)\,\rho_{1}(\Vec{r}_2)\,,
\label{contraction}
\eeq
where $\rho_2(\Vec{r}_1,\Vec{r}_2)$  is the two-body density matrix
\beq
\rho_2(\Vec{r}_1,\Vec{r}_2)=\int \prod_{j=3}^A d\Vec{r}_j
\,\left|\psi_0(\Vec{r}_1,\Vec{r}_2 ...,\Vec{r}_A)
\right|^2 \delta\left(\sum (\Vec{r}_j)\right)\,.
\label{twobody}
\eeq
The  one- and two-body density matrices appearing in Eq. (\ref{psiquadro}) are
normalized according to
\beq
\label{nordens}
\int d\Vec{r} \, \rho_1(\Vec{r})\,=
\,\int d\Vec{r}_1\,d\Vec{r}_2\,\rho_2(\Vec{r}_1,\Vec{r}_2)\,=\,1
\eeq
and satisfy the following sequential conditions:
\beq
\int\,d\Vec{r}_j\,\rho_1(\Vec{r}_i,\Vec{r}_j)\,\rho_1(\Vec{r}_j,\Vec{r}_k)
\,=\,\rho_1(\Vec{r}_i,\Vec{r}_k)
\label{twoseque}
\eeq
\beq
\int\,d\Vec{r}_2\,\rho_2(\Vec{r}_1,\Vec{r}_2)\,=\,\rho_1(\Vec{r}_1)\,,
\label{oneseque}
\eeq
which leads to
\beq
\label{propdelta}
\int d\Vec{r}_1\,\Delta(\Vec{r}_1,\Vec{r}_2)\,=\
\,\int d\Vec{r}_2\,\Delta(\Vec{r}_1,\Vec{r}_2)\,=0;
\eeq
In  Eq. (\ref{psiquadro}) only   unlinked contractions have to be considered, and
the  higher order terms include unlinked products of 3, 4, {\it etc}. two-body contractions,
unlinked products of three-body contractions, describing three-nucleon correlations, and so on.
When all terms up to A-body correlations are written down explicitly, an identity is obtained.

The common  approximation in  Glauber type calculations
consists in disregarding all terms of Eq. (\ref{psiquadro}) but the first one. In this
case the very well known expression
for the total Glauber profile is given by
\beq
\label{quattordici}
\Gamma_{00}^G(\Vec{b}_n)= 1 - \left[1\,-\,\int d\Vec{r}_1\,\rho_{1}(\Vec{r}_1)
      \,\Gamma(\Vec{b}_n-\Vec{s}_1)\right]^A\,.
\eeq
By taking into account  two-body correlations, i.e. all unlinked  products of two-body
contractions in Eq. (\ref{psiquadro}),
one obtains \cite{moniz,rafa}
\beqy
\Gamma_{00}^G(\Vec{b}_n)&=& 1 - \left[1\,-\,\int d\Vec{r}_1\,\rho_{1}(\Vec{r}_1)
      \,\Gamma(\Vec{b}_n-\Vec{s}_1)\right]^A\, \times\nonumber\\
     &&\hspace{-1cm}\times\, \sum_{m=0}^{\left[\frac{A}{2} \,or\, \frac{A-1}{2}\right]}
\frac{A!}{(A-2\,m)!m!}\left\{\frac{1}{2}
      \,\frac{\int d\Vec{r}_1 d\Vec{r}_2\,\Delta(\Vec{r}_1,\Vec{r}_2)\,
      \Gamma(\Vec{b}_n-\Vec{s}_1)\,\Gamma(\Vec{b}_n-\Vec{s}_2)}
      {\left[1-\int d\Vec{r}_1\,\rho_{1}\Vec{r}_1)
      \,\Gamma(\Vec{b}_n-\Vec{s}_1)\right]^2}\right\}^m
\label{exact}
\eeqy
which in
the optical limit (A $>>$ 1) becomes
\beq
\Gamma_{00}^G(\Vec{b}_n)\simeq
1-e^{\displaystyle{-A\int d\Vec{r}_1
      \,\rho_{1}(\Vec{r}_1)\,\Gamma(\Vec{b}_n-\Vec{s}_1)+
\frac{A^2}{2}{\int d\Vec{r}_1 d\Vec{r}_2\,\Delta(\Vec{r}_1,\Vec{r}_2)\,
      \Gamma(\Vec{b}_n-\Vec{s}_1)\,\Gamma(\Vec{b}_n-\Vec{s}_2)}}}.
\label{optical}
\eeq
As for the inelastic shadowing contribution (\ref{Fdelta}),  it can be reduced to an
expression depending upon the total nucleon and diffractive cross sections
$\sigma _N^{tot}$ and $\sigma _r^{tot}$ respectively.
\beqy
\sigma_A^{IS}&=&\frac{4 \pi}{k}Im \Gamma_{00}^{IS}(0)\,=\,2
\int d\Vec{b}_n\,\Gamma_{00}^{IS}(\Vec{b}_n)\,=\nonumber\\
&=&-4 \pi^2A^2\int d\Vec{b}_n\,dz_1\,dz_2\,\rho (\Vec{b}_n, z_1)\rho (\Vec{b}_n, z_2)
\int dM_X^2\,\frac{d^2\sigma}{d^{q_{T}^2}dM_{X}^2}\Big|_{q_{T}=0}\,\cdot\nonumber\\
&&\cdot\,\left[\,\Theta (z_1-z_2)\,e^{\displaystyle{i(p_{lab}-p_{m})(z_2-z_1)}}
\,\cdot\right.\nonumber\\
&&\left.\cdot\,e^{-A}\int_{z_2}^{z_1}dz^\prime\,\rho(\Vec{b}_n, z^\prime)\,\frac{\sigma_{r}}{2}
\,\cdot\,e^{-A}\int_{-\infty}^{z_2}dz^\prime\,\rho(\Vec{b}_n, z^\prime)\,\frac{\sigma_{T}}{2}
\cdot\,e^{-A}\int_{Z_1}^{+\infty}dz^\prime\,\rho(\Vec{b}_n, z^\prime)\,\frac{\sigma_{T}}{2}
\,+\right.\nonumber\\
&&\left.+\,\Theta (z_2-z_1)e^{\displaystyle{i(p_{lab}-p_{m})(z_1-z_2)}}\,\cdot\right.\nonumber\\
&&\left.\cdot\,e^{-A}\int_{z_2}^{z_1}dz^\prime\,\rho(\Vec{b}_n, z^\prime)\,\frac{\sigma_{r}}{2}
\,\cdot
e^{-A}\int_{-\infty}^{z_1}dz^\prime\,\rho(\Vec{b}_n, z^\prime)\,\frac{\sigma_{T}}{2}
\cdot e^{-A}\int_{Z_2}^{+\infty}dz^\prime\,\rho(\Vec{b}_n, z^\prime)\,\frac{\sigma_{T}}{2}\,\right]
\,+\nonumber\\&&-\,\int dz_1 dz_2\,e^{\displaystyle{i\,q_L(z_1-z_2)}}
\,\Delta(\Vec{b}_n,z_1,\Vec{b}_n,z_2).
\eeqy
Within the approximation $\sigma _N^{tot}= \sigma _r^{tot}$ and disregarding correlations
($ \Delta(\Vec{b}_n,z_1,\Vec{b}_n,z_2)=0$)
the well-known Karmanov-Kondratyuk \cite{karmanov}  expression is obtained
\beq
\sigma_A^{IS}\,=\,-4 \pi A^2\int d\Vec{b}_n\, \int
\frac{d^2\sigma}{d^2q_{T}\,dM_{X}^2}\,\Big|_{q_{T}=0} dM_{X}^2
e^{\displaystyle{-\frac{\sigma_{T}}{2}\,T(b)}}|F(q_{L}, z)|^2\,,
\label{KK}
\eeq
where $q_L$ depends upon $M_x^2$.

\begin{figure}[!htp]
\centerline{\centerline{\epsfig{file=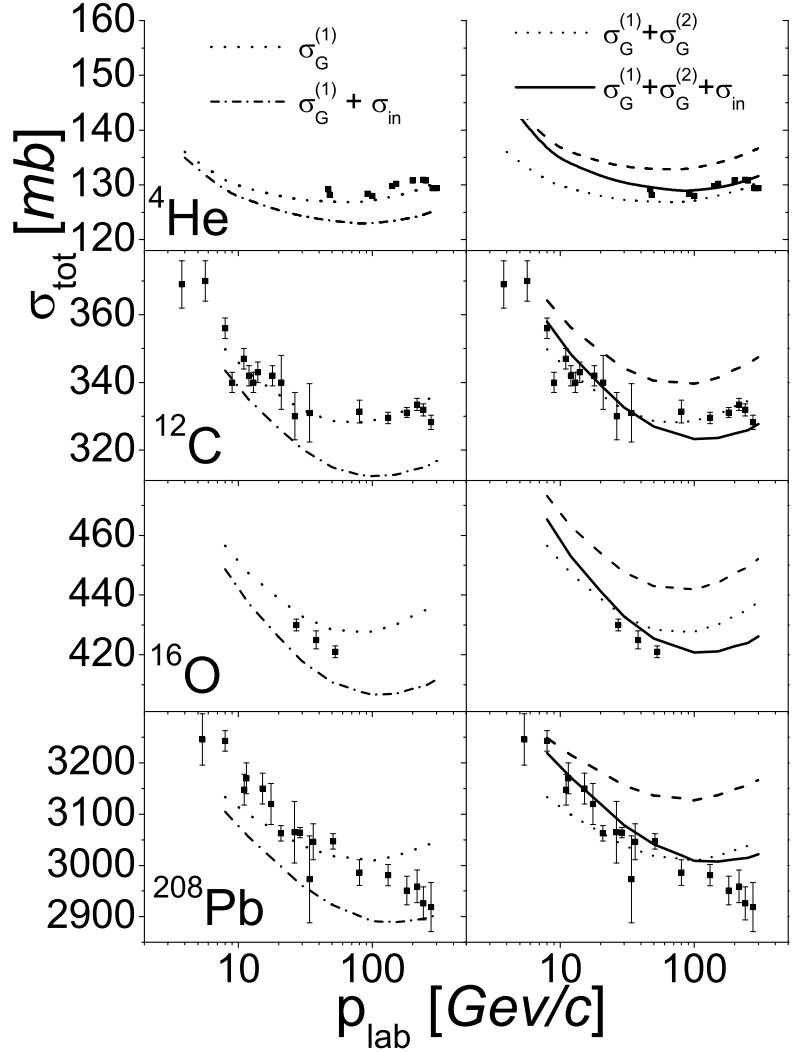,width=12cm}}}
\caption{The neutron-nucleus total cross section, for $^{4}He$, $^{12}C$, $^{16}O$ and
  $^{208}Pb$. \textit{Left panel}: the result without the inclusion of correlations;
  \textit{dotted curves}: one-body (first) term of Eq. (\ref{optical});
  \textit{dot-dashed curves}:
  one-body term plus inelastic shadowing effects (Eq. \ref{KK}). \textit{Right panel}:
  results with the inclusion of correlation.
  \textit{Dotted curves}: one-body (first) term of Eq. (\ref{optical});
  \textit{dashed curves}: one-body term plus two-nucleon correlations of Eq. (\ref{optical});
  \textit{solid curves}:  one-body term plus two-nucleon correlations of Eq. (\ref{optical})
  plus inelastic shadowing effects (Eq. \ref{KK}). Experimental data from
  \cite{moniz,data02}}
\label{Fig2}
\end{figure}

In our calculations we have used both Eqs. (\ref{optical}) and (\ref{KK}). We have checked
that the optical limit for $A=16$ is valid within $1 \%$, whereas  correlations produce very
tiny effects on $\sigma_A^{IS}$. The ingredients of our  calculations were  as follows:

\begin{enumerate}

\item The density matrices have been  obtained by a linked cluster expansion
 for the one- and two-body density
operators expectation value, evaluated over a fully-correlated wave function
\cite{alv01} obtained variationally with the Argonne $V8^\prime$ interaction \cite{vuotto}.
The one-body density has been obtained by integrating the two-body density. Let us
stress that, unlike previous calculations, our two-body contractions (Eq. (\ref{contraction}))
exactly satisfy the condition given by Eq. (\ref{propdelta});
\item the Glauber profile function is of the usual form
\beq
\Gamma(\Vec{b}_n)\,=\,\frac{\sigma_{tot}}{4\,\pi\,b_0}\,(1-i \alpha)
\,e^{\displaystyle{-\Vec{b}_n^2 / b_0}}
\eeq
with the energy-dependent parameters taken from \cite{param_glau};
\item the parameters for the inelastic shadowing were taken from \cite{murthy}.
\end{enumerate}
The results of calculations for $^4He$, $^{12}C$ , $^{16}O$ and $^{208}Pb$ are presented
in Fig. \ref{Fig2}. The left panel shows the results obtained without correlations,
whereas  the effects of correlations are presented in the right panel.
The results presented in Fig. \ref{Fig2} deserve the following comments:
\begin{enumerate}
\item it can be seen that correlations  increase the total cross section by
about $10\%$,  i.e. they {\it decrease} the nuclear transparency, worsening
the agreement with the experimental data when  only Glauber shadowing is considered;
the inclusion of inelastic shadowing  brings  back theoretical calculations in good
agreement with experimental data;
\item in the case of $^4He$ we have calculated the cross section to all orders of correlations
using the exact wave function of Ref. \cite{hiko}; it turns out that three- and four-nucleon
correlations produce negligible effects on the total cross section;
\item
the effect of correlations is of the same order as the one from inelastic shadowing.
\end{enumerate}

It should be pointed out that the contribution to the optical phase shift
 (the second term in the exponent of
 Eq. (\ref{optical}))
is always negative; the black disk limit of our approach is satisfied.

\begin{figure}[!hp]
\centerline{
  \epsfysize=0.8\textwidth\epsfbox{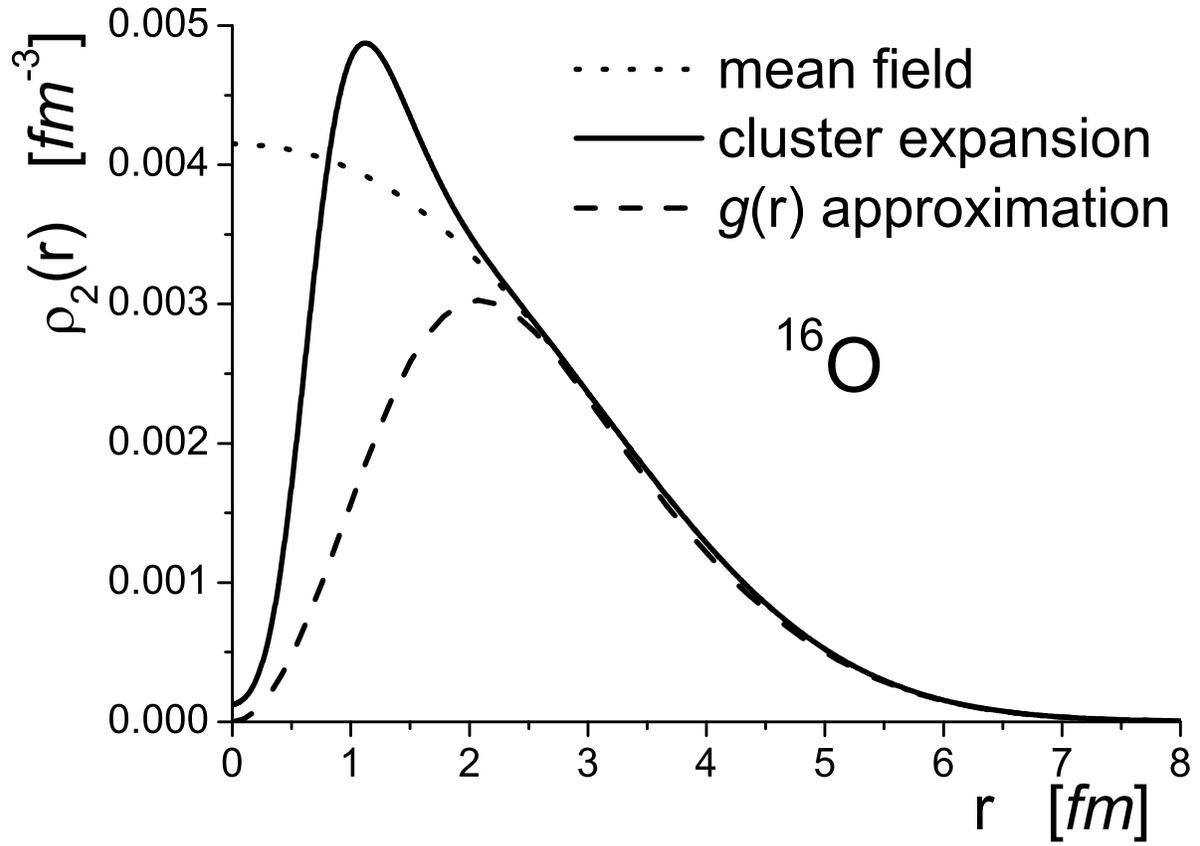}}
\caption{The two-body density matrix of $^{16}O$ within various approximations:
  the mean field approximation (\textit{dots}), the approximation of Ref.
  \cite{moniz} (\textit{dashes}), the realistic calculation of the present
  work (\textit{full}).}
\label{Fig3}
\end{figure}

\begin{figure}[!hp]
\centerline{
  \epsfysize=0.8\textwidth\epsfbox{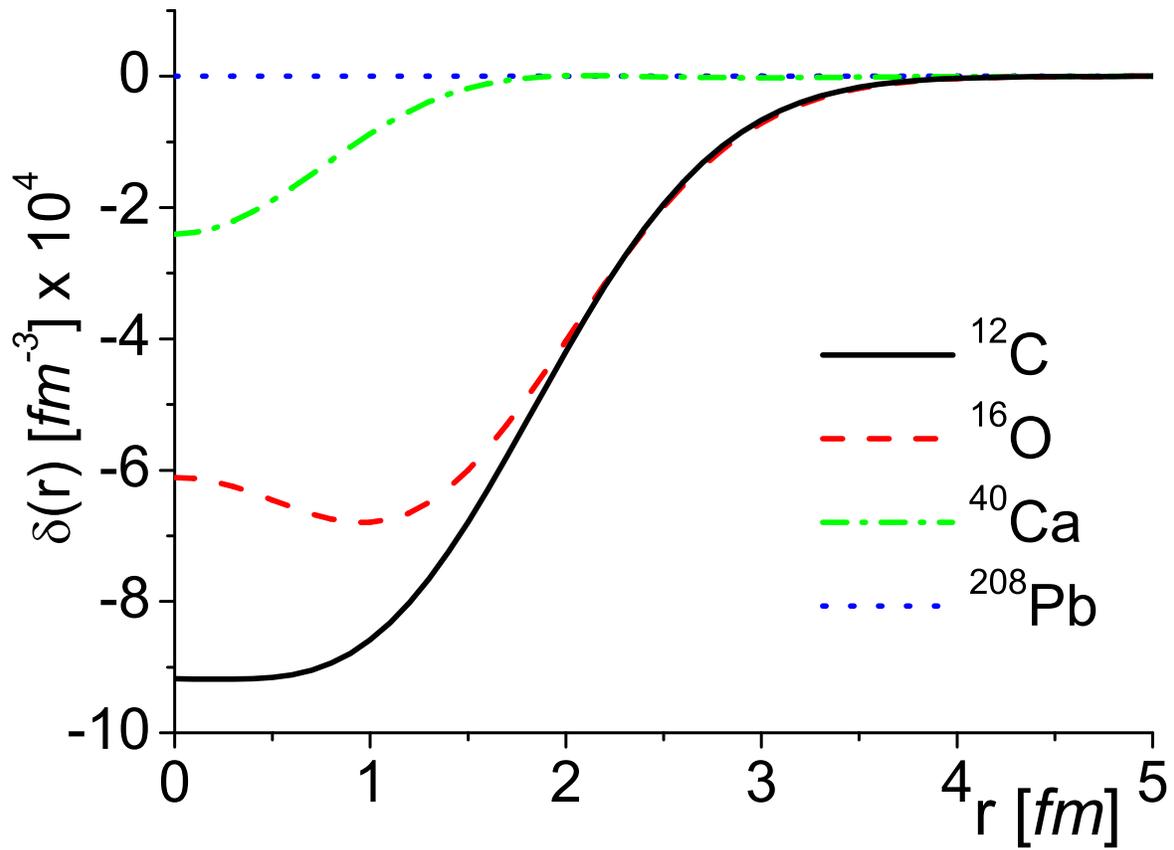}}
\caption{The quantity (\ref{violation}), calculated within the approximation
  $\rho_2(\Vec{r}_1,\Vec{r}_2)\,=\,\rho_1(\Vec{r}_1)\,\rho_1(\Vec{r}_2)\,g(r_{12})$,
  for various nuclei.}
\label{Fig4}
\end{figure}

\begin{figure}[!hp]
\centerline{
  \epsfysize=0.8\textwidth\epsfbox{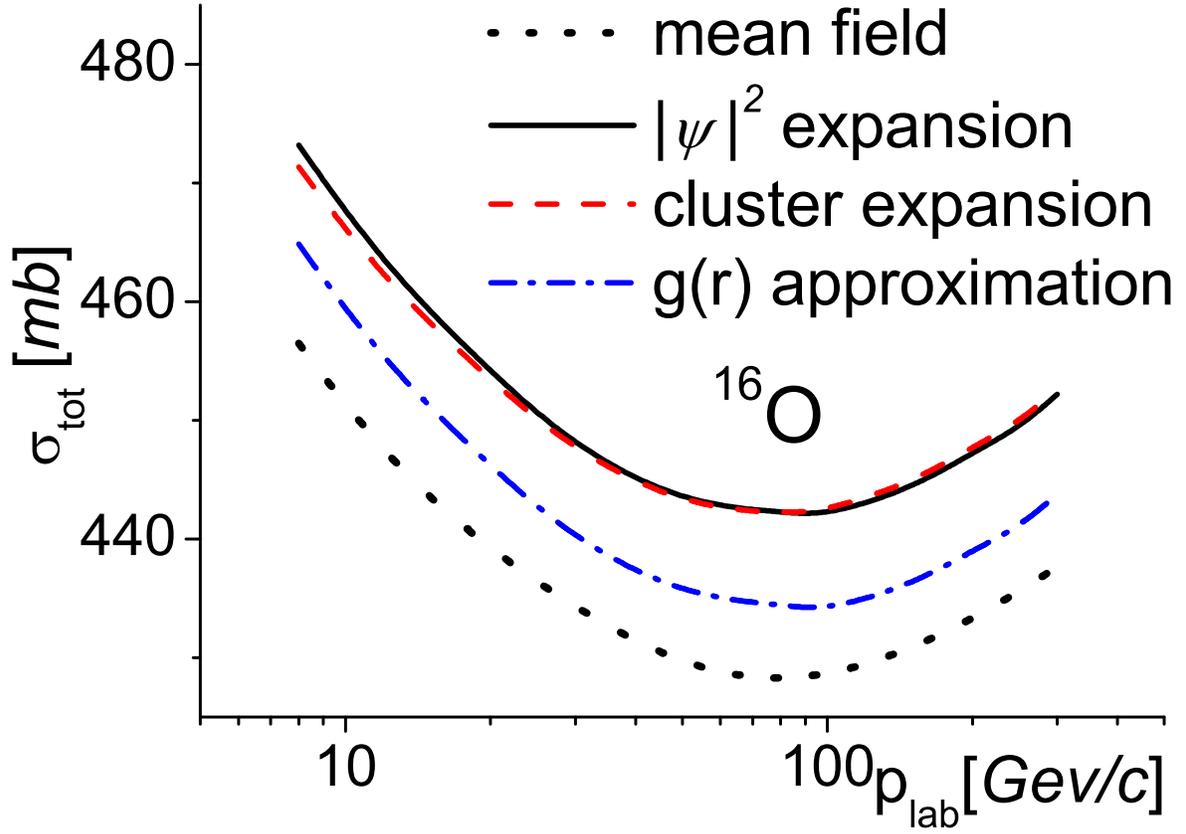}}
\caption{Results of calculations for $\sigma_{tot}$ using the $\rho_D(\Vec{r},\Vec{r}^\prime)$
  of Refs. \cite{cladan,croazia} (\textit{red, dashed curve}) as compared to the result
  corresponding to the wave function expansion of Ref. \cite{foldy}
  (\textit{black, solid curve}); the \textit{blue, dash-dotted curve} is obtained with
  the correlation model of Ref. \cite{moniz}
  the \textit{black, dotted curve} is the usual mean-field result.}
\label{Fig5}
\end{figure}

In Fig. \ref{Fig3} the difference between the two-body density
calculated within the mean field, the cluster expansion and the following approximation
for the two-body density
\beq
\label{propgi}
\rho_2(\Vec{r}_1,\Vec{r}_2)\,=\,\rho_1(\Vec{r}_1)\,\rho_1(\Vec{r}_2)
\,g(|\Vec{r}_1-\Vec{r}_2|)\,
\eeq
frequently used in case of complex nuclei (see \textit{e.g.} Ref. \cite{moniz}),
is exhibited. The curves represent the quantity
\beq
\rho_2(r)\,=\,\int d\Vec{R}\,\rho_2\left(\Vec{r}_1=\Vec{R}+\frac{1}{2}\Vec{r}\,,\,
\Vec{r}_2=\Vec{R}-\frac{1}{2}\Vec{r}\right)\,.
\label{approxi}
\eeq
It can be seen that the two-body density of Eq. (\ref{approxi}) represents a poor representation
of the realistic one. We have checked to what extent the sequential relation (\ref{propdelta})
is violated by the approximate  two-body density matrix.
 The amount of violation of the integral in Eq. (\ref{propdelta}) can be checked
by calculating the quantity
\beq
\label{violation}
\delta(r_2)\,=\,\int d\Vec{r}_1\,\Delta(\Vec{r}_1,\Vec{r}_2)\,;
\eeq
which  is shown in Fig. \ref{Fig4} for various nuclei. Fig. \ref{Fig5} shows the effect
of violation of the sequential relation on the total cross section.

%-------------------------------------------------------------------------------SECTION
\section{A cluster expansion approach to the total cross section}

We have developed a cluster-expansion \cite{alv01} formulation for $\sigma^{tot}_{nA}$
based upon the one-body \textit{distorted} density matrix of Ref. \cite{cladan},
obtained taking into account two-nucleon correlations at first order of the cluster
expansion and Glauber rescatterings at all orders. The zeroth-order approximation
(\ie{ }with no correlation effects) is the same as Eq. (\ref{quattordici});
correlations can be included
with the first term of the wave function expansion of Eq. (\ref{psiquadro}),
by replacing the one-body densities appearing in such a term with the distorted
one-body density proposed in Ref. \cite{cladan}, in such a way one obtains
contributions representing the interaction of the incident nucleon with the particles
involved in each of the diagrams contributing to the distorted density (namely, the
\textit{shell model}, \textit{hole} and \textit{spectator} diagrams). The final
expression for the total cross section reads as follows:
\beq
\sigma^{tot}_{nA}\,=\,\sigma^{tot}_{SM}+\,\Delta\sigma^{tot}_{H}+\,\Delta\sigma^{tot}_{S}\,,
\eeq
in which the shell model ($SM$), hole ($H$) and spectator $(S)$ contributions are
as follows:
\beqy
\sigma^{tot}_{SM}&=&2\int d\Vec{b}_n\,\Big[1\,-\,\left(1\,-\,\frac{4}{A}\,
\int d\Vec{r}_1\,\rho_o(\Vec{r}_1)\,\Gamma(\Vec{b}_n,\Vec{b}_1)\right)^A\Big]\nonumber\\
%------------
\Delta\sigma^{tot}_{H}&=&2\int d\Vec{b}_n\,\left(1\,-\,\frac{4}{A}\,
\int d\Vec{r}_1\,\rho_o(\Vec{r}_1)\,\Gamma(\Vec{b}_n,\Vec{b}_1)\right)^A\,\times\nonumber\\
&&\hspace{0cm}\times\,\frac{4}{A}\,\int d\Vec{r}_1 d\Vec{r}_2\,
\Big(4\,H^D_{12}\,\rho_o(\Vec{r}_1)\,\rho_o(\Vec{r}_2)\,-\,
H^E_{12}\,\left|\rho_o(\Vec{r}_1,\Vec{r}_2)\right|^2\Big)\,
\left[G_1(\Vec{b}_n)\,G_2(\Vec{b}_n)\,-\,1\right]\nonumber\\
%------------
\label{cluExpCla}
\Delta\sigma^{tot}_{S}&=&-2\int d\Vec{b}_n\,\left(1\,-\,\frac{4}{A}\,
\int d\Vec{r}_1\,\rho_o(\Vec{r}_1)\,\Gamma(\Vec{b}_n,\Vec{b}_1)\right)^A\,\times\nonumber\\
&&\times\,\frac{4}{A}\,\int d\Vec{r}_1\,d\Vec{r}_2\,d\Vec{r}_3\,
\rho_o(\Vec{r}_1,\Vec{r}_2)\Big(4\,H^D_{23}\,\rho_o(\Vec{r}_2,\Vec{r}_1)
\,\rho_o(\Vec{3})\,+\nonumber\\
&&\hspace{2cm}-\,H^E_{23}\,\rho_o(\Vec{r}_2,\Vec{r}_3)\,\rho_o(\Vec{r}_3,\Vec{r}_1)\Big)\,
G_1(\Vec{b}_n)\,G_2(\Vec{b}_n)\,\Gamma_1(\Vec{b}_n)\,.
\eeqy
Note that in the final calculations the shell model term has been exponentiated
as in Eq. (\ref{optical}).

The results of calculations for $^{16}O$ obtained with Eq. (\ref{cluExpCla}) are compared
in Fig.  \ref{Fig5}, with the results predicted, by Eqs. (\ref{psiquadro}) and
(\ref{propgi}), respectively.
It can be seen that the expansion based on the distorted density of Ref. \cite{cladan}
is in perfect agreement with the one of the approach based on the expansion of the wave
function, Eq. (\ref{optical}), despite the different class of diagrams appearing in each
contribution.
%, while the approximation of Eq. (\ref{propgi}) underestimates $\sigma_{tot}$
%by a few percent.

It should be stressed that the expansion used for the distorted one-body density
can be used to calculate, as in Ref. \cite{cladan}, the total transparency in
$A(e,e^\prime p)X$ experiments and distorted momentum distributions, as in Refs.
(\cite{croazia,qmbt}) with the full correlated wave function for complex nuclei.
Calculations of nuclear and color transparencies in $(e,e^\prime p)$ and $(p,2p)$
are in progress and will be reported elsewhere.

%-------------------------------------------------------------------------------SECTION
\section{Summary and conclusions}

We have developed  a method which can be used to calculate scattering processes at medium
and high energies within a realistic and parameter-free  description of nuclear structure.
Our calculations followed  the following strategy:
\begin{itemize}
\item[i)] the values of the parameters pertaining to the correlation functions and
the mean field wave functions, have been  obtained from the calculation of the ground-state
energy, radius and density of the nucleus using realistic nucleon-nucleon interactions;
\item[ii)] using these parameters we have calculated the total neutron-Nucleus cross section
taking rigorously into account two-nucleon
correlations within the expansion of the exact wave function \ref{psiquadro}. We have
also adopted a cluster expansion procedure \cite{cladan} obtaining essentially
the same results. This gives us confidence that the treatment of correlations is
model independent to a large extent.

The method we have developed appears to be a very effective, transparent and parameter-free
one and the main results we have obtained are:
\item[i)] the effects generated by
two-nucleon correlations (\ie{ }by those parts of the wave function expansion
(\ref{psiquadro}) which contain two-body contractions), are of the same order as Gribov
inelastic shadowing; this, in our opinion, points to the necessity of an analysis
of the accuracy of the common approximation used in medium-high energy hadronic scattering
processes consisting in disregarding all terms of the expansion (\ref{psiquadro}) but
the first one;
\item[ii)] correlations due to three and higher order contractions appear
to produce only negligible effects on the total cross section.
\end{itemize}

\section{Acknowledgements}

We are indebted to Daniele Treleani, Boris Kopeliovich and Nikolai Nikolaev
for many illuminating discussions.

%%%%%%%%%%%%%%%%%%%%%%%%%%%%%%%%%%%%%%%%%%%%%%%%%%%%%%%%%%%%%%%%%%%%%%%%%%%
%%%%%%%%%%%%%%%%%%%%%%%%%%%%%%%%%%%%%%%%%%%%%%%%%%%%%%%%%%%%%%%%%%%%%%%%%%%
%%%%%%%%%%%%%%%%%%%%%%%%%%%%%%%%%%%%%%%%%%%%%%%%%%%%%%%%%%%%%%%%%%%%%%%%%%%
%%%%%%%%%%%%%%%%%%%%%%%%%%%%%%%%%%%%%%%%%%%%%%%%%%%%%%%%%%%%%%%%%%%%%%%%%%%

%%%%%%%%%%%%%%%%%%%%%%%%%%%%%%%%%%%%%%%%%%%%%%%%%%%%%%%%%%%%%%%%%%%%% Fig 1
\newpage
%%%%%%%%%%%%%%%%%%%%%%%%%%%%%%%%%%%%%%%%%%%%%%%%%%%%%%%%%%%%%%%%%%%%% Fig 2
\newpage
%%%%%%%%%%%%%%%%%%%%%%%%%%%%%%%%%%%%%%%%%%%%%%%%%%%%%%%%%%%%%%%%%%%%% Fig 3
\newpage
%%%%%%%%%%%%%%%%%%%%%%%%%%%%%%%%%%%%%%%%%%%%%%%%%%%%%%%%%%%%%%%%%%%%% Fig 4
\newpage
%%%%%%%%%%%%%%%%%%%%%%%%%%%%%%%%%%%%%%%%%%%%%%%%%%%%%%%%%%%%%%%%%%%%% Fig 5
\newpage
%---------------------------------------------------------------------
\end{document}